\documentclass[%
 reprint,
 amsmath,amssymb,
 aps,
]{revtex4-2}

\usepackage{graphicx}
\usepackage{dcolumn}
\usepackage{bm}

\begin{document}

\preprint{APS/123-QED}

\title{Long-Lived Acoustic Phonon and Carrier Dynamics in III-V Adiabatic Cavities}

\author{Muhammad Hanif$^1$}
\author{Milos Dubajic$^2$}%


%
\author{Sujakala J. Sreerag$^3$}%
\author{Rajeev N. Kini$^3$}%
\author{Gavin J. Conibeer$^1$}
\author{Michael P. Nielsen$^1$}
\author{Stephen P. Bremner$^1$}
\email{stephen.bremner@unsw.edu.au}
\affiliation{$^1$
 School of Photovoltaics and Renewable Energy Engineering, UNSW Sydney, Australia
}
\affiliation{$^2$Department of Physics, University of Cambridge, Cambridge United Kingdom
}%

\affiliation{$^3$Indian Institute of Science Education and Research, Thiruvananthapuram, India}%
\date{\today}

\begin{abstract}
We provide evidence of strongly confined coherent acoustic phonons inside high quality factor phononic cavities that exhibit tailored phonon potentials. 
Using GaAs/AlAs quasiperiodic superlattices, we have realized functional phonon potentials by adiabatically changing the layer thicknesses along the growth direction. 
Room temperature ultrafast vibrational spectroscopy reveals discrete phonon modes with frequencies in the range of $\approx 96-101$ GHz.
Additionally, we confirm that phononic cavities significantly retard the energy loss rate of the photoexcited carriers, as evidenced by time-resolved photoluminescence measurements. 
These results highlight the potential of concurrently engineering optoelectronic and phononic properties for a range of novel applications.

\end{abstract}%


\maketitle

\section{\label{sec:level1}Introduction }

Advances in nanofabrication accompanied by improved characterization techniques have shown the potential for engineering control and developing an understanding of the dynamics of phonons in solids at an atomic scale \cite{PhysRevB.97.165412}.
Analogous to electronic properties, phononic properties are described by a dispersion relation $\omega(q)$, where $\omega$ and $q$ are the angular frequency and the wave vector, respectively. 
Phonons are quasiparticles with an energy $E\textsubscript{p} = \hbar \omega$, momentum $p = \hbar q$ and obey Bose-Einstein statistics. 
On the basis of the onset and characteristic frequency, phonons can be categorized into optical and acoustic phonons.
Optical phonons, which originate from out-of-phase atomic movement, significantly impact the optical properties of semiconductors, while acoustic phonons, which result from in-phase atomic displacement, are the predominant heat carriers in solid-state materials \cite{lanzillotti2007phonon}.
Thus understanding and control of phononic dynamics are vital for applications, including energy conversion \cite{pmid:22996556}, optomechanics \cite{aspelmeyer2014cavity}, nanoelectronic \cite{pop2010energy}, and biomedical diagnostics \cite{doi:10.1063/1.2988470}.

In comparison to engineering photonic and electronic properties at the nanoscale, control over phononic properties has proven extra challenging due to confinement effects, the atomistic nature of matter, and efficient phononic interactions with electrons and photons \cite{volz2016nanophononics}. 
Superlattice structures enable the investigation of confinement for electrons, photons, and phonons, since the wavelengths of visible light and low-frequency (GHz) phonon modes are comparable. Such confinement can be achieved by designing the physical properties through tailoring the layer thicknesses in the superlattice structure. Superlattices made of III-V binaries, for example, offer pristine material quality and precise control over layer thicknesses and atomically sharp interfaces through well developed epitaxial deposition techniques like molecular beam epitaxy. These have been employed in various nanoacoustic devices, including phonon mirrors, filters, and resonant cavities \cite{doi:10.1063/1.3295701,doi:10.1063/1.5000805}.
Inspired by their optical counterparts, phononic Distributed Bragg Reflector (DBR) based isolated and coupled cavities have been studied extensively to develop an understanding of simultaneous confinement of photons and phonons, efficient optophononic coupling, manipulate lattice thermal conductivity, and achieve higher phonon control \cite{balandin1998significant,trigo2002confinement,lanzillotti2007phonon,pmid:16108421, loo2010quantum,lanzillotti2015towards,anguiano2017micropillar,esmann2019brillouin,sandeep2019phoniton,ortiz2021coherent}. The results reported therein are promising and call for strategies to achieve a higher phononic control to develop practical devices.

In a superlattice, modifications to the phonon dispersion occur through periodic changes in the crystal structure, leading to the formation of minibands and minigaps attributed to quantum effects and interference within the superlattice.
The tailoring of phonon potentials, defined as the central frequency of varying minigap in localised acoustic phonon dispersion in a supperlattice structure, can be a vital instrument to spatially drive or confine phonons \cite{ortiz2019phonon}.
Such phononic potentials have been proposed by defining a bi-layer unit-cell as the fundamental building block of a superlattice. The thickness of the constituent layers within the unit-cell gradually changes along the growth direction to locally engineer the acoustic phonon dispersion, while maintaining a constant overall thickness of the unit-cell \cite{doi:10.1063/1.3295701}. A generalized model has recently been developed to design high Q-factor phononic cavities with the desired effective phonon potential based on adiabatic tuning of the superlattice periodicity \cite{ortiz2019phonon}. 
Nevertheless, an experimental demonstration of coherent acoustic phonon (CAP) properties in such tailored phonon potentials is limited. 

At room temperature, phonons in semiconductors exist across a broad spectral range and actively participate in the determination of the electronic transport properties.
Characterizing a system with such broad-spectrum phonon excitations is demanding, both per se and due to interaction with photons and photogenerated electrons \cite{li2012colloquium}. 
Usually, Raman spectroscopy is employed to study phononic properties in materials that rely on the inelastic scattering of incident laser light by phonons \cite{trigo2002confinement,esmann2018topological,colvard1980observation}. 
Furthermore, Raman shifts originating from sub-100 GHz acoustic phonons ($<3 cm^{-1}$), which usually exist in phononic nanocrystals and acoustic nanocavities, are difficult to resolve due to the associated instrument response \cite{ng2022excitation}.
However, time-domain ultrafast vibrational spectroscopy (UVS) has shown its potential to excite coherent phonons in the sub-THz range and record photoinduced acoustic echoes in differential transmission and reflectivity \cite{thomsen1984coherent,thouin2019phonon}. 
Recently, UVS has been used to resolve 23.6 GHz acoustic phonons in bulk CdSe \cite{wu2015ultrafast}. 
In addition, topological nanophononic states around 150 GHz in sandwiched DBR structures have been successfully captured in the UVS experiment
\cite{esmann2018topological,arregui2019coherent}.

This work presents phonon dynamics in high-Q acoustic phononic cavities with tailored phonon potential, realized using GaAs/AlAs superlattices, and characterized via UVS. Additionally, the photoexcited carrier cooling energy loss rate was probed using time-resolved photoluminescence at room temperature.

\section{\label{sec:level2}Designing Acoustic Cavities with Tailored Phonon Potentials}

Phonon confinement, usually achieved through phononic cavities, is an important component of phononic engineering.  The strength of confinement depends on the physical shape of the cavity and the acoustic impedance mismatch between the cavity constituents. 
A material system with established optoelectronic and mechanical properties, such as GaAs/AlAs, is a suitable candidate for archetypal phononic cavity devices.
To study the effect of different phonon potentials on phononic confinement, we have designed phononic cavities exhibiting 1D parabolic and quartic potentials along the growth axis. 
The samples are grown via molecular beam epitaxy (MBE) on lattice-matched GaAs substrate (growth details can be found in Supplementary Information S-3).

\begin{figure}[ht!]
\begin{center}
\includegraphics[width=.85\columnwidth]{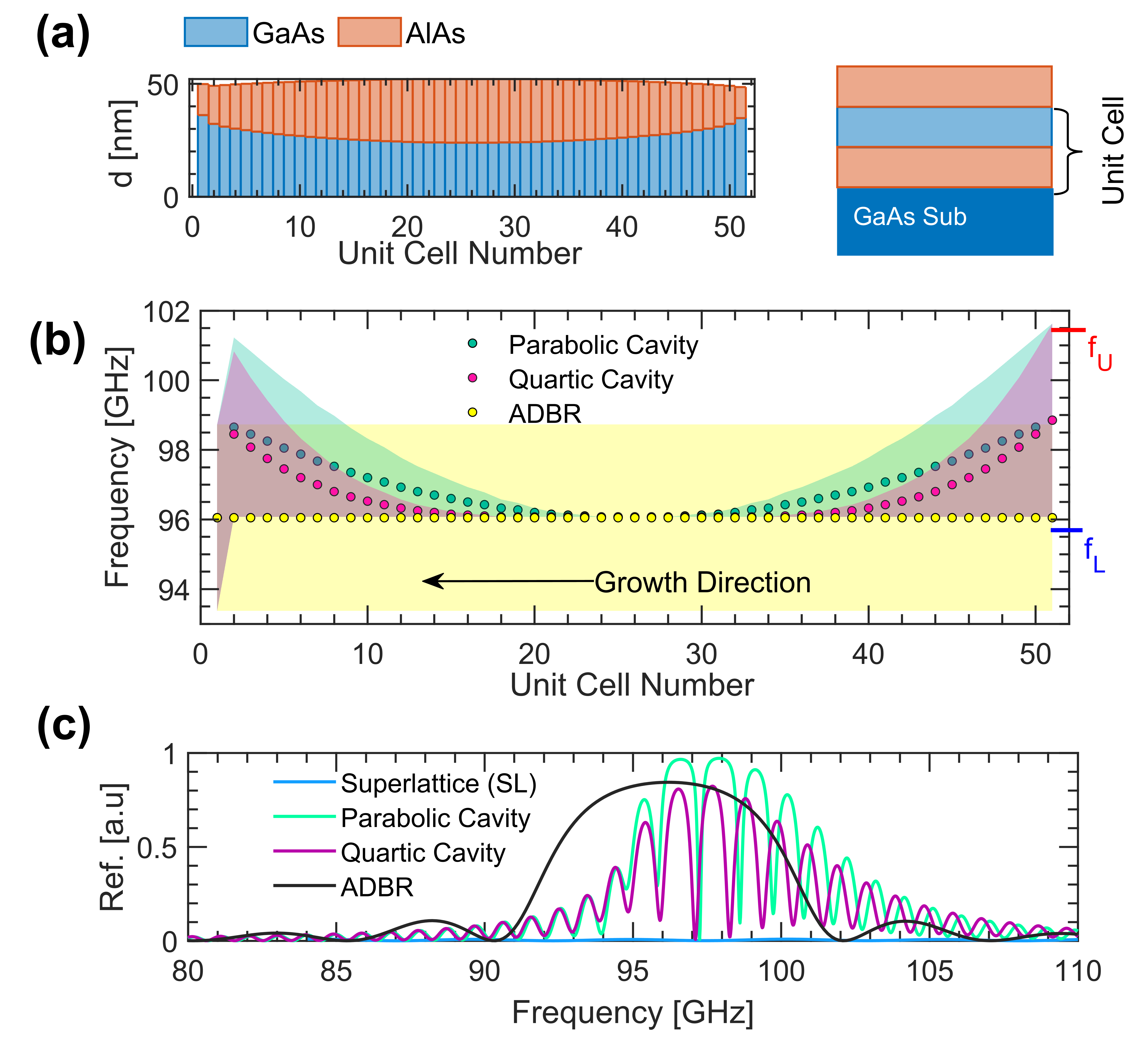}
\caption{{Phononic cavities with tailored acoustic phonon potentials modeled using Equation 1-3, (a) Illustrates the construction of a unit cell (right), and varying GaAs/AlAs layer thicknesses within each unit cell of constant extent, $d_{cell}$, to realize functional potential. (b) Shows parabolic and quartic phonon potentials (markers) and the respective longitudinal acoustic phonon minigap in each unit cell (shaded). The yellow shaded area represents the phonon gap of the acoustic distributed Bragg reflector (ADBR) sample.  
(c) Acoustic reflectivity of designed samples calculated by the transfer matrix method. A superlattice, which has no acoustic phonon gap in its phonon dispersion, is also included for comparison of reflectivity. 
{\label{cavity_design}}%
}}
\end{center}
\end{figure}

Phonon potentials are realized by gradually varying the thicknesses of the GaAs/AlAs constituent bilayers, where each bilayer can be described as a unit cell, as shown in Figure {\ref{cavity_design}}a \cite{ortiz2019phonon}.
The minigaps at the center of the Brillouin zone in longitudinal acoustic phonon (LAP) dispersion are opened or closed by changing GaAs/AlAs thicknesses while keeping the acoustic path length in a unit cell unchanged.
A unit cell of constant thickness also facilitates adiabatic tuning of the minigap, which helps reduce interfacial phonon scattering. Adiabatic tuning refers to the gradual change in acoustic phonon minigap energy and position in successive unit cells of the phononic cavity.

Furthermore, the stacking periodicity of GaAs/AlAs affects the optoelectronic properties \cite{jiang2020effects}. Here, we constrain the constituent layers to be more than 20 monolayers in thickness, to ensure that the direct optical band gap of the cavity sample is matched to that of bulk GaAs (1.42 eV) in order to focus on the phononic properties of the structure without introducing a significant carrier confinement. 

To design the phononic cavities with functional potentials, we begin with the parameterized phonon dispersion relation in a periodic superlattice \cite{ortiz2019phonon}; 

\begin{equation}
\cos\left(\text{qd}\right)=\cos\left(\frac{\omega D}{f_{des}}\right)-\frac{\epsilon^{2}}{2}\sin\left(\frac{\omega rD}{f_{des}}\right)\sin\left(\frac{\omega\left(1-r\right)D}{f_{des}}\right) 
\label{eq1}
\end{equation}

\begin{equation}
\epsilon=\frac{Z_1-Z_2}{\sqrt{Z_1Z_2}}
\label{eq2}
 \end{equation}

Where $q$ is the phonon momentum, $d={d_1+d_2}$ is the thickness of the unit cell, $\omega$ is the angular frequency of the phonon. The subscripts $1$ and $2$ refer to GaAs and AlAs, respectively, $Z_{1,2}=\nu_{1,2}\rho_{1,2}$ is the acoustic impedance, $\nu$ and $\rho$ is the acoustic velocity and density of the respective material. $D=D_1+D_2$ is the total acoustic path length normalized to the wavelengths in the respective materials at the design frequency, $f_{des}$, with $D_{1,2}=\frac{d_{1,2}f_{des}}{\nu_{1,2}}$. 
The parameter $r=D_2/D$ gives the contribution of the AlAs layer to the total length of the acoustic path of the unit cell and $f_{des}$ is the designed confining frequency.  

To experimentally investigate nonequilibrium acoustic phonon dynamics in 1D adiabatic acoustic phononic cavities, we designed phononic cavities consisting of 51 cells with effective 1D parabolic and quartic potentials given by ~\(f(x)=f_{0}{x}^{2}\) and~\(f(x)=f_{0}{x}^{4}\), respectively, where \(x\)~is the growth axis of the superlattice and is defined as the number of unit cells from the front surface, as shown in Figure 1.
In order to choose the minigap in the dispersion of acoustic phonons at the center of the Brillouin zone $q$ is set to $0$ in Equation \ref{eq1} and $D = 1$ to preserve the length of the acoustic path in each unit cell. The upper and lower cut-off points of the phonon confinement range are defined as $f_U$ and $f_L$, respectively. The value of $f_L$ is set to $96$ GHz. With \(f_L\) fixed, the upper stopband~\((f_U)\) is varied smoothly in each unit cell
along the growth direction to achieve a continuous phonon potential such
that\(f(x)=f_L\) for $x=1$, and \(f(x)={\{[{f_{U}-f_{L}](x)}^{2,4}+f_{L}\}}\) for $x = 2\rightarrow N$, determines the local phonon gap in each unit cell and the center of the phonon minigap is given by~\(f(x)\), illustrated in Figure \ref{cavity_design}b. With these substitutions, Equation \ref{eq1} reduces to 
\begin{equation}
1=\cos(\frac{{2\pi f}_{L}}{f(x)})-\frac{\epsilon^{2}}{2}\sin(\frac{{2\pi f}_{L}r}{f(x)})\sin(\frac{{2\pi f}_{L}(1-r)}{f\left(x\right)})
 \end{equation}

Equation 3 can be numerically solved for $r$, and therefore the thicknesses of the constituent layers of the unit cellfound, as illustrated in Figure 1a. Notably, the minigap of the first unit cell is redshifted to align its central frequency with the vertex of the cavity potential, which lowers the reflectivity of the front (top) face of the sample, somewhat analogous to a laser cavity. This makes phonon excitation and detection from the top of the structure possible and is called the probe cell.

To conduct a comparative study, on phonon dynamics and photoexcited energy loss dynamics, four samples are considered, which include a parabolic and quartic phonon cavities where constituent GaAs/AlAs layers in parabolic and quartic cavities vary from 24.9 to 37.6 nm for GaAs and 14.4 to 29.4 nm for AlAs. An acoustic distributed Bragg reflector (ADBR) sample is realized by fixing $r = 0.24$ and grown as a $20$ period GaAs/AlAs (37.6 nm/14.4 nm) stack. A superlattice sample (SL) without an acoustic minigap is also designed by setting $r = 0.5$ for comparison. The SL also has 20 periods GaAs/AlAs (24.9 nm/29.4 nm). All the samples except for the gapless SL are designed to operate at 96 GHz, as confirmed by reflectivity calculations depicted in Figure \ref{cavity_design}c using the acoustic transfer matrix method.

\section{Non-equilibrium Phonon Dynamics}
A significant component of understanding non-equilibrium dynamics is the phonon lifetime, which is defined as the time a phonon takes to lose coherence, scatter or attenuate. 
The lifetime of coherent acoustic phonons (CAPs) determines the damping time of the phonon energy and the attenuation factor \cite{cuffe2013lifetimes}. 
An adequate lifetime and coherence of phonons allow robust control and help preserve and store information \cite{liu2022effect}.
 
We employed Ultrafast Vibrational Spectroscopy (UVS) to study phonon dynamics. 
In the UVS technique, the photons from each incoming ultrashort pulse are coherently scattered into lower-frequency photons, producing phase-coherent phonons $(\omega, \pm{k})$, which form a standing wave in the cavity. 
Vibrational oscillations are detected by coherent scattering of variably delayed probe pulses whose reflection or transmission is modulated by the presence of phonons \cite{yan1985impulsive,pmid:15865401}.

Ultrashort pump pulses trigger interactions among incident photons, carriers, and cavity phonons, which depend on the optoelectronic and phononic properties of the samples. 
Phonons are coherently excited after ultrafast pulses impinge on the sample through a series of competing processes. The initially excited non-equilibrium carriers undergo fast energy and momentum relaxation to achieve a Fermi-Dirac distribution by loosing their excess energy to phonons \cite{othonos1998probing}. Coherent phonons are also generated by pump pulse through optomechanical coupling, in which case the radiation pressure is transferred to the lattice atoms, resulting in phononic oscillations \cite{de2021strong}. Alternatively,
direct optoacoustic coupling can also lead to the generation of CAPs
\cite{hess1993laser}. In a solid-state semiconductor excited above the bandgap, some combination of all these processes will result in the final coherent phonon packet developing.

\begin{figure*}[ht!]
\includegraphics[ width=1.8\columnwidth]{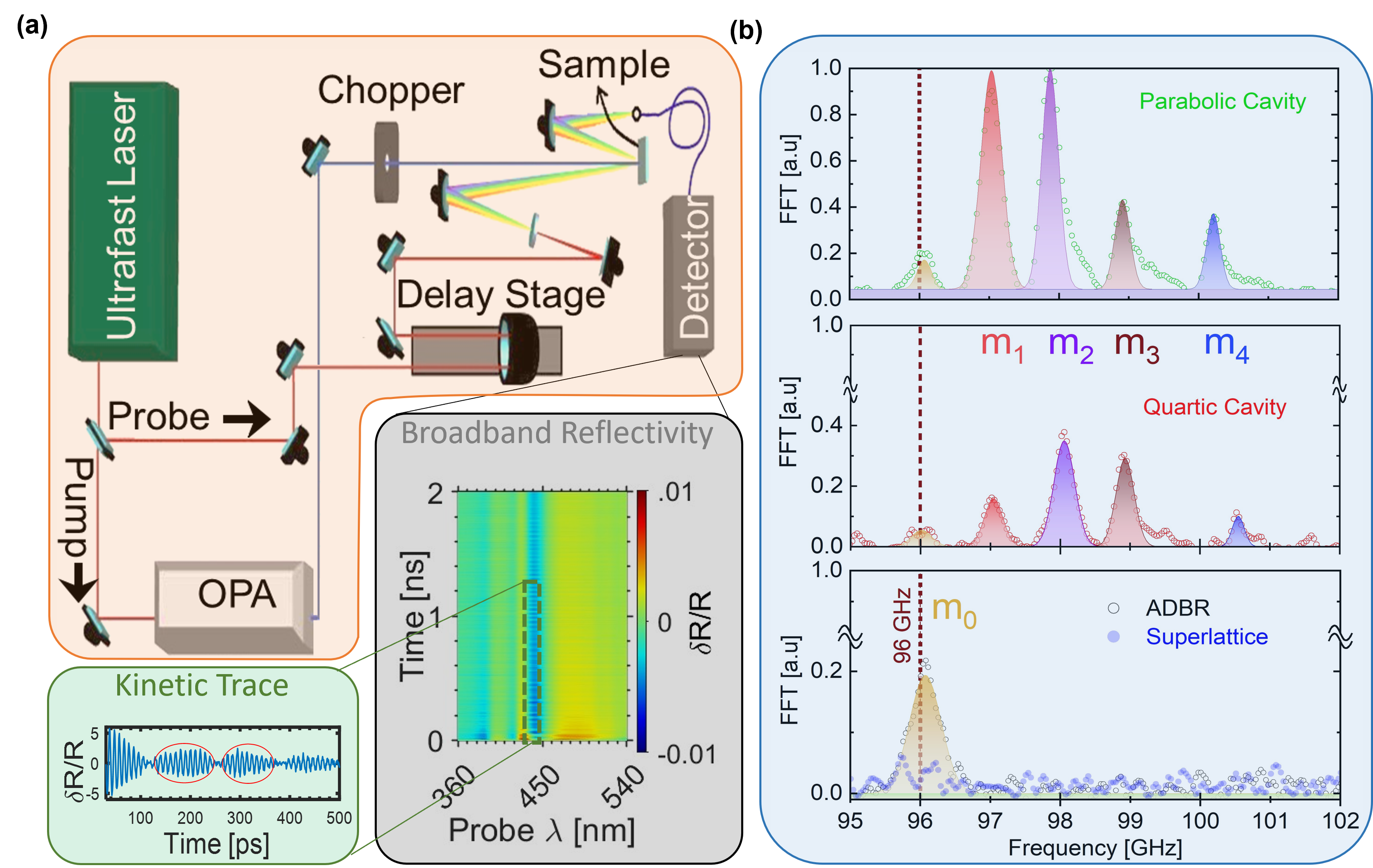}
\caption{{(a) Broadband ultrafast vibrational spectroscopy setup. The insets, gray and green, show an example of measured broadband differential reflectivity, and the electronic background filtered kinetic trace. (b) Phonon dynamics extracted from the measured differential reflectivity of the parabolic cavity, quartic cavity, ADBR, and superlattice sample by taking an FFT of spectrally integrated kinetics from  $\lambda_{prob} = 415-450$ nm to improve signal to noise ratio. The shaded peaks represent the Gaussian fitting. (See the Supplementary Information S-1 for details on phonon information retrieval from UVS reflectivity data). 
{\label{UVS schematic_FFT}}%
}}
\end{figure*}

The penetration depth and fluence of the pump pulse determine the excitation strength of the cavity modes. 
Shorter wavelength pump pulses excite only surface modes. Thus, the penetration depth of the pump pulse is chosen to be larger than the thickness of the probe cell for stronger optomechanical coupling.

For above bandgap energies, increasing pump fluence enhances indirect phonon excitations by absorbing energy from excited electrons. 
This process adds an exponentially decaying electronic background to the transient response of the cavity, that must be removed in order to analyse the phononic response. An example of the electronic background removal process as well as transient reflectivity spectra before background removal are shown in the Supplementary Material.

The selection of the probe spectrum is also critical for capture of the dynamics of cavity CAPs that evanescently leak through the phonon potential.
The penetration depth of the probe wavelength was chosen to be half of the thickness of the probe unit cell (50nm in our case). This makes the UV-vis probe spectra suitable for detecting CAPs in the designed
samples. 

Here, we excited the samples with 610 nm pump pulses of $2.71$ mJ/cm$^2$ and probed them using a broadband supercontinuum probe from 350 to 600 nm. 
The probe beam is p-polarized (with respect to the sample) and incident at 45\textsuperscript{o} to improve the probe sensitivity to the phonons. However, the normal incidence pump beam is depolarized to ensure that all phonon modes are excited. 
The measurement setup is shown in Figure {\ref{UVS schematic_FFT}}a. 
Cavity phonons perturb the incident probe, which is sensitive to refractive index changes caused by modulation in the mean atomic positions and their square displacement. The detector is capable of capturing broadband reflectivity, shown in the inset of Figure \ref{UVS schematic_FFT}a. 
Then a fast Fourier transform (FFT) of an electronic background filtered and spectrally integrated from $\lambda_{probe} = 415-450nm$  kinetics is taken and shown in Figure \ref{UVS schematic_FFT}b corresponding to the parabolic, quartic and ADBR sample. The spectral integration of kinetics is employed to improve the signal to noise ratio. 
Four confined CAP modes, labeled m\textsubscript{1-4}, are found in the parabolic and quartic cavities and are well correlated with the calculated acoustic reflectivity shown in Figure {\ref{cavity_design}}c. In Figure {\ref{UVS schematic_FFT}}b, phonon dynamics of the gapless superlattice (SL) sample (filled markers) and ADBR sample with single confined mode m\textsubscript{0} are shown. 
In the FFT spectrum of the parabolic and quartic potential cavities, Figure {\ref{UVS schematic_FFT}}b, we observe a phonon mode at $f_L= 96.07$ GHz, marking the bottom of the designed phonon potentials. 
This is confirmed by the FFT spectrum of the ADBR sample, which is designed to exhibit a central frequency of 96 GHz, shown in Figure {\ref{UVS schematic_FFT}}b, and labeled as~\(m_0\) (shaded yellow).

The peaks associated with CAP modes in the baseline-corrected FFT spectra are fitted with Gaussian functions to estimate the full width half maximum (FWHM) that is related to lifetime $(\tau)$ as $\tau = 1/(2\pi FWHM)$ and quality factor (Q-factor) as Q-factor = $f_m/FWHM$ of the individual modes.
The Q-factor associated with $m_2$, the strongest CAP mode, of the parabolic and quartic phonon cavity is $1.81\pm 0.05 \times 10^3$ and $1.76 \pm 0.05 \times 10^3$, respectively. Significantly, lower FFT signal strength and marginally lower Q-factor in the quartic cavity can be attributed to higher interfacial losses emanating from sharper phonon potential transitions. The Q-factor associated with $m_0$, acoustic phonon mode of ADBR sample, is $1.16 \pm 0.04 \times 10^3$. The FFT spectra of the SL sample show no CAP mode since it is designed to have a zero acoustic phonon minigap and thus no phononic reflectivity. 

The effective cavity width $(ECW)$, defined as the spatial separation between the opposing reflective interfaces of the cavity potential, puts a lower bound on the cavity width. The frequency-dependent ECW in parabolic and quartic phononic cavities helps to establish multiple resonant phonon modes. 
The results indicate an increase in the total travel length of the phonon width from $m_1$ to $m_4$ for the parabolic and quartic cavities. An increase in the length of the travel of the phonon and the finite phonon potential causes the FFT amplitude, and therefore the energy stored in each mode, to deteriorate monotonically from $m_2$ to $m_4$ \cite{PhysRevLett.110.037403}. 

\section{\label{sec4} Temporal dynamics of cavity phonons }

\begin{figure}[ht!]
\begin{center}
\includegraphics[width=\columnwidth]{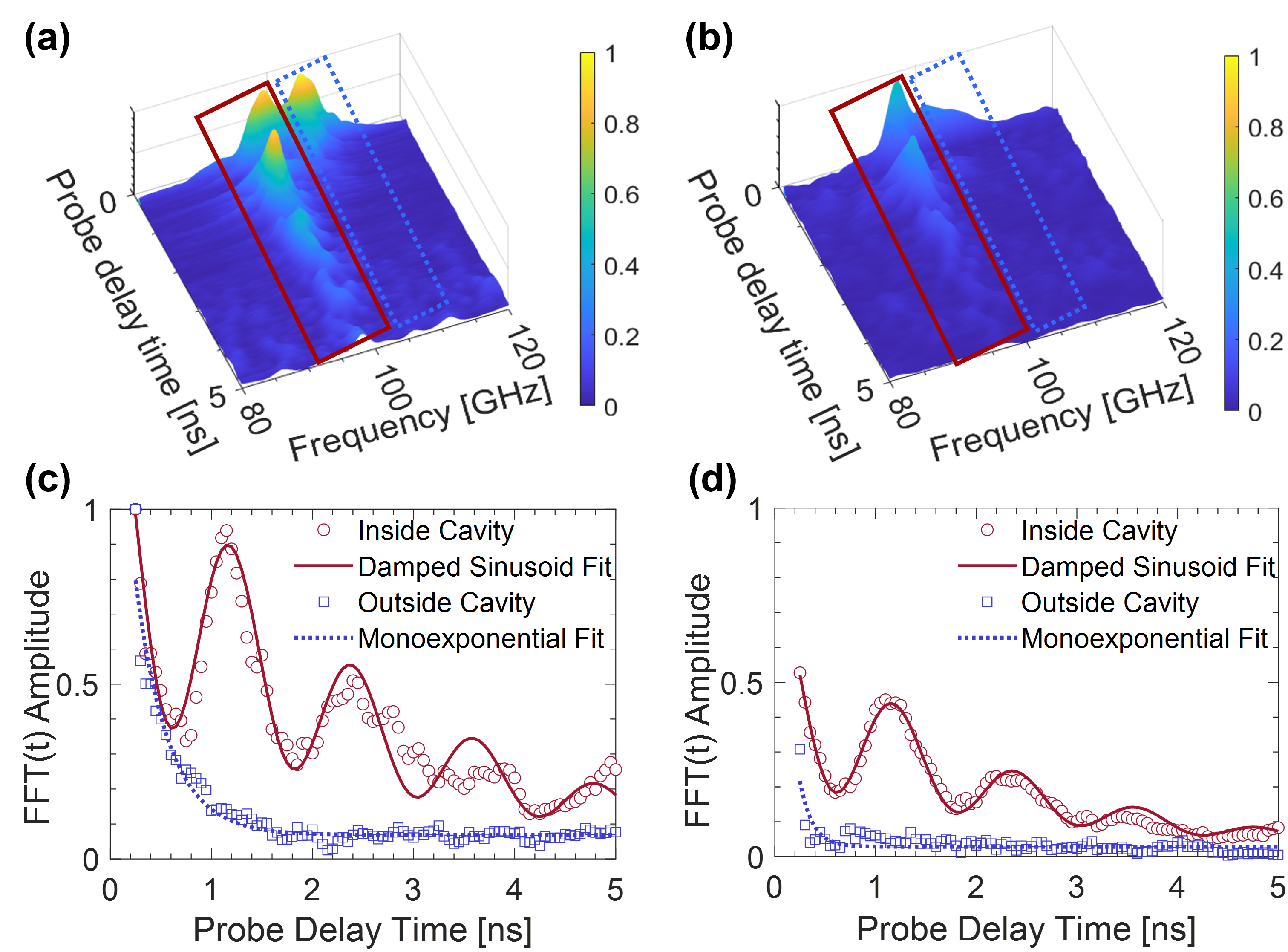}
\caption{{Windowed Fourier transform of parabolic cavity (a), and quartic cavity (b).  (c) and (d) Show the respective FFT peak intensities as a function of time inside (red) and outside (blue) of the cavities. The solid and dashed curves in (c) and (d) are fits. 
{\label{WFFT}}%
}}
\end{center}
\end{figure}

The temporal phonon dynamics of cavity samples were elucidated by taking a sliding-window Fourier transform of the measured differential reflectivity \cite{lanzillotti2015towards}. The window Fourier transform as a function of time is shown for the parabolic and quartic cavities in Figure \ref{WFFT} (a) and (b), respectively. 
Relinquishing the spectral resolution, the time evolution of the cavity phonons can be captured.

In order to resolve the dynamics of confined vs bulk phonons, the time window (t\textsubscript{step}) was chosen to be 0.5 ns, with overlap between adjacent time windows. The results are shown Figure \ref{WFFT}a and Figure \ref{WFFT}b for parabolic and quartic cavities, respectively. The colormaps in Figure \ref{WFFT} suggests that the ultrafast pump ($\lambda$ = 610 nm) excites the phonon in a broad spectral range at an early time. Phonon perturbations with frequencies that are favored by the cavity potential design, marked with solid rectangles, survive longer than phononic oscillations with frequencies outside the designed phonon confining potential, as highlighted with dashed rectangles in Figure \ref{WFFT}a-b. 
The cavity oscillations are characterized by plotting the peak intensity of the FFT(t) signal in the Figure \ref{WFFT} c-d for the parabolic and quartic cavities, respectively. The FFT(t) peak data in the phonon confining frequency range has been fitted with an empirical function function to estimate the frequency and decay constant: $\Phi_{fit}=A*cos^2( 2\pi f t) \times \exp{(-t/\tau)}$.
Here, A is the initial amplitude of the FFT(t), $f$ is the oscillation frequency, and $\tau$ is the coherent phonon lifetime.  Fitting the oscillations in the measured data results in a frequency ($f$) of 0.83 $\pm$ 0.06 GHz for both cavities, corresponding to a period (T) of 1.2 ns. The coherent phonon lifetime ($\tau$) extracted from the decay has values of 2.22 $\pm$ 0.01 ns and 1.82 $\pm$ 0.01 ns for parabolic and quartic cavities, respectively. Outside the cavity potential, acoustic phonon population decay faster with a life time of 0.33 $\pm$ 0.03 ns for the parabolic and 0.14 $\pm$ 0.06 ns for the quartic cavity.

Since the samples are made up of approximately 52.3\% GaAs (1363.5 nm) and 47.7\% AlAs (1242 nm), the average bulk longitudinal acoustic speed through the stack would be $\approx$ 5199.5 m/s \cite{tamura1999phonon}. Total thickness of multilayers in the cavity samples are $\approx$ 2.6$\mu$m. This results in an alternative estimation of the cavity round trip time of 1 ns, which closely matches to the 1.2 ns extracted from the fitting, potentially indicating slightly different acoustic speed at these frequencies compared to the average. This suggests that the oscillatory behavior inside the cavity most likely arise from back and forth reflections of the coherent phonons. For phonons modes with frequencies not targeted by the cavity design, the phonon amplitude diminishes exponentially as expected, with results presented in blue (symbols) and exponential fits (dashed line) in Figure \ref{WFFT}c-d for the parabolic and quartic cavity, respectively.  

\section{\label{sec5} Excess Energy Loss rate of Photoexcited charge carriers }

To study the effect of phononic manipulation on the optoelectronic properties of the structures, we performed time-resolved photoluminescence (TRPL) measurements with 532 nm excitation at a fluence of $2.17$ mJ/cm$^2$. The laser pulses of a width of 10 ps at a 1 MHz repetition rate were focused onto the sample using a 0.3NA (15x) silver reflective objective. The emission spectrum was fed to a monochromator for spectral filtering and finally collected by a time-correlated single photon counting (TCSPC) system, allowing us to collect spectrally and temporally resolved emission data. 

\begin{figure}[ht!]
\begin{center}
\includegraphics[width=\columnwidth]{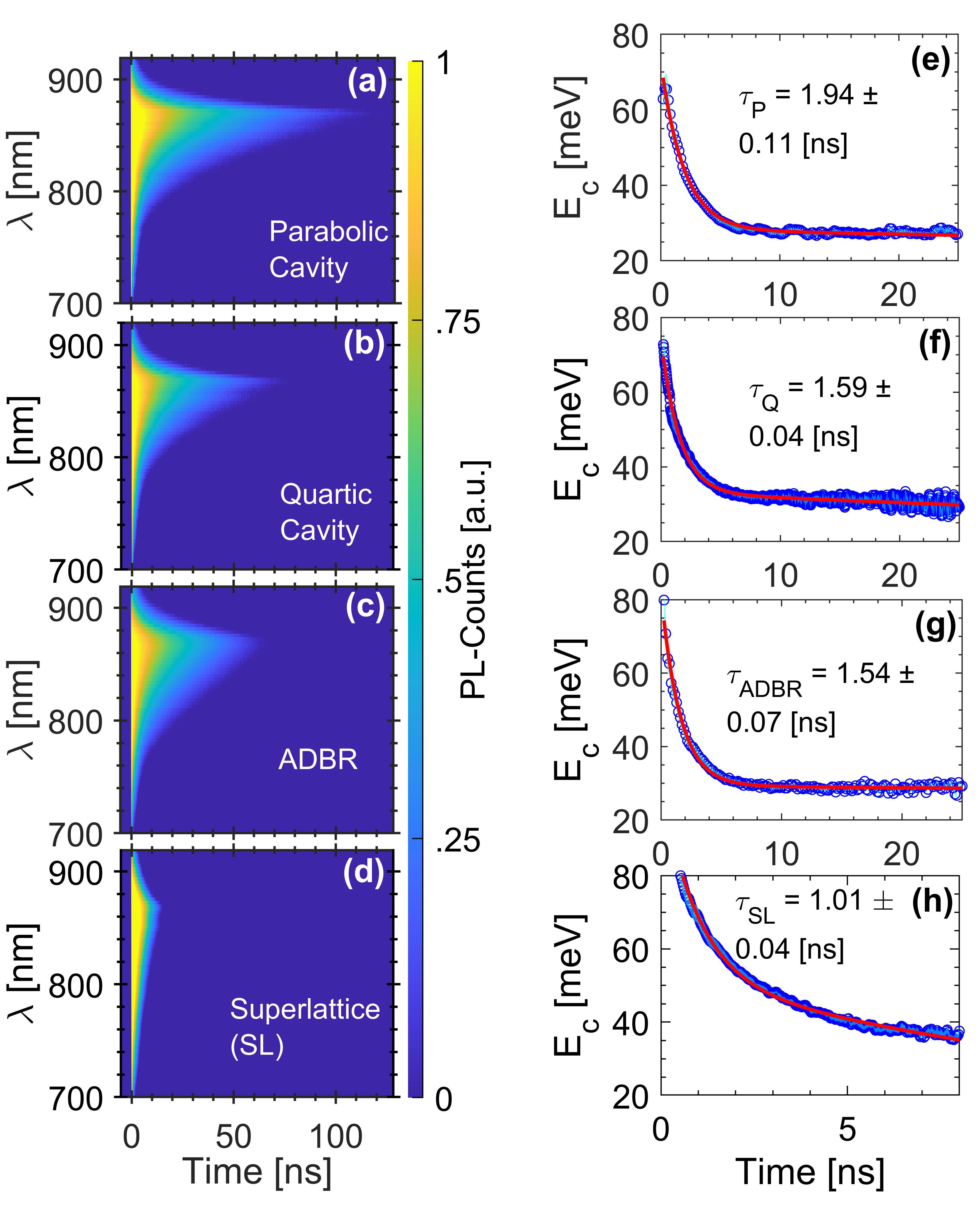}
\caption{{2D TRPL colormaps of  parabolic, quartic, ADBR, and SL sample are shown in (a) to (d), respectively. PL emission in all the samples peaks at 864 nm but survives proportionally to the phonon confinement strength. The solid (red) curves therein are biexponential function fit to estimate carrier energy loss rate. Only the longer decay constants are printed as $\tau_P$, $\tau_Q$, $\tau_{ADBR}$, and $\tau_{SL}$ for comparison. 
{\label{435055}}%
}}
\end{center}
\end{figure}

2D-TRPL colormaps of parabolic, quartic, ADBR, and superlattice samples are shown in Figure {\ref{435055}}(a-d), respectively. It is evident from the PL peak energy that the electronic band edge has not been changed due to the phononic cavity design and matches well with bulk GaAs at room temperature at 1.42 eV. However, unlike bulk electronic band structures, light hole (lh) and heavy hole (hh) bands are distinct to give a step-like emission spectrum. 

Figure 4(a-d) indicates that PL emission lasts significantly longer in the parabolic cavity and dies quickly as the phonon confinement strength decreases, from parabolic to quartic to ADBR to the SL sample.  
Time dependent photoexcited carrier energies are estimated by full spectrum PL fitting with a multipeak function which takes into account average kinetic energy of photoexcited carriers $E_c(t)$ along with aforementioned electronic transitions. 
This method for $E_c$ estimation is similar to the well-known carrier temperature estimation from the linear fitting of the high-energy tail on a logarithmic scale \cite{gibelli2016accurate}. However, hot carrier temperature estimations assume that the lattice remains at or near room temperature in equilibrium whereas in our system we are purposely confining a range of acoustic phonons which are the thermal energy carriers whose population dynamically changes over time. As such, we cannot ascribed that the photoexcited carriers are necessarily at a higher temperature than the lattice and only assign them an average kinetic energy compared to room temperature. On a longer time scale $(>30 ns)$, the average carrier energy ($\approx$ 26 meV) reduces to the equivalent room temperature (301.3 K) which points to a reliable $E_c$ estimation \cite{beeby2012condensed}. 
More details on the $E_c$ estimation are given in the Supplementary Information S-2.


The energy loss rate is finally estimated from a biexponential fitting of time dependent carrier energies with results presented in Figure {\ref{435055}}(a-d).
The short decay components found from fitting of the data are $1.94\pm 0.11$ ns, $1.59\pm 0.04$ ns,  $1.54\pm 0.07$ ns, and $1.01\pm 0.04$ ns for the parabolic, quartic, ADBR, and SL sample, respectively. This is in good agreement with the coherent phonon lifetimes extracted from the temporal dynamics of the cavity phonons, indicating that phonon confinement has enabled the carriers to hold onto their energy longer. This is also indicated by the decay times tracking the quality factor of the cavities, with higher quality factor cavities having a slower carrier energy loss rate. The long time component, on the order of >100 ns, represents the sample cooling after the excitation pulse and is not attributed to the phonon confinement.
It is important to note that the widening of the high-energy photoluminescence is not solely linked to an elevation in carrier temperature. Additional factors, like state filling where carriers populate higher-order confined states in the quantum well (QW), and broadening caused by phonons, contribute to the expansion of the linewidth in the PL spectrum \cite{esmaielpour2017effect}. 
Interestingly, the phonon minigap in the ADBR sample ($5$ GHz$ = 0.02$ meV) is also much smaller than $18$ meV (half of the optical phonon energy $36$ meV in GaAs at $\Gamma$-point) required to inhibit Klemens decay \cite{konig2010hot}, which may indicate the potential for the the phonon bottleneck effect (PBE) \cite{conibeer2017towards,esmaielpour2021hot} in these samples.

A retarded excess energy loss rate in the phononic cavities compared to the SL sample show a strong correlation with phonon confinement \cite{brockmann1994hot,makhfudz2022coherent}. However, the intricacy of scattering mechanisms in superlattices, as reported in the literature \cite{PhysRevMaterials.6.L061001}, driving photoexcited carriers energy loss requires further investigations. For example, temperature, excitation power, and wavelength dependent PL/TRPL coordinated with UVS measurements can selectively permit or inhibit different scattering mechanisms to provide an unambiguous understanding.

\section{Conclusion}
 Using III-V binary superlattices, we have demonstrated high quality factor (Q-factor) acoustic cavities with a range of tailored 1D phonon potentials. 
 Ultrafast vibrational spectroscopy unambiguously confirms the presence of 
 coherent acoustic phonons confinement. 
 The parabolic cavity showed the longest coherent phonon decay time of 2.2 ns and the highest Q-factor of $\approx 2 \times 10^3$ for an individual phonon mode. The achieved Q-factor is comparable to hBN-MoS\textsubscript{2} cavities \cite{zalalutdinov2021acoustic} and DBR coupled quantum dot cavities \cite{loo2010quantum}.  
Additionally, time resolved photoluminescence  measurements confirmed a significantly slower excess energy loss rate of photoexcited carriers in phononic cavities compared with DBR sample.
In addition to obtaining control over realizing 1D phonon potentials, the observation of a slowed energy loss rate of photoexcided charge carriers in phononic cavities presented in this work provides a promising basis to develop applications including phononic crystals, waveguides, and thermoelcetrics.
\\
\\
\\
Contribution and acknowledgment: S.P.B. proposed the project. M.H. and M.D. did device modeling with advice from S.J.S and R.N.K. S.P.B fabricated the samples. M.H., M.D., and M.P.N. designed the experiments, and M.H. and M.D. performed the measurements. M.H. did the data analysis. M.H., M.P.N. and S.P.B. wrote the manuscript. All the authors discussed the results and contributed to the final manuscript. S.P.B. and M.H. acknowledge support of ANFF-UNSW for sample growth via MBE. M.P.N. acknowledges the support of the UNSW Scientia Program and the Australian Research Council (ARC) Centre of Excellence in Exciton Science (project No: CE17100026).


\bibliography{PhononCavityDraft}

\end{document}